\documentclass[fleqn,usenatbib]{mnras} %
\usepackage[multidot]{grffile}

\usepackage[utf8]{inputenc}
\usepackage[T1]{fontenc}
\usepackage{ae,aecompl}

\usepackage{graphicx}	
\usepackage{amsmath}	
\usepackage{amssymb}	

\usepackage{wasysym}
\usepackage{courier}
\usepackage{mathrsfs}
\usepackage{threeparttable}
\usepackage{verbatim}
\usepackage{comment}
\usepackage[british]{babel}
\usepackage{color}
\usepackage{booktabs}
\usepackage{tabularx}

\newcommand{\beq}{\begin{equation}}
\newcommand{\eeq}{\end{equation}}
\newcommand{\kepler}[0]{\emph{Kepler}}
\newcommand{\galaxia}[0]{\emph{Galaxia}}

\newcommand{\teff}[0]{$T_\text{eff}$}
\newcommand{\rsolar}[0]{$R_{\astrosun}$}

\newcommand{\Dnu}[0]{$\Delta\nu$}

\newcommand{\numax}[0]{$\nu_{\rm max}$}

\newcommand{\fdnu}[0]{$f_{\Delta\nu}$}

\defcitealias{yuj++2018-16000-rg}{SYD18}
\defcitealias{pinsonneault++2018-apokasc}{APK18}
\defcitealias{sharma++2019-k2-hermes-age-metallicity-thick-disc}{G19}

\newcommand{\yu}[0]{\citetalias{yuj++2018-16000-rg}}
\newcommand{\apk}[0]{\citetalias{pinsonneault++2018-apokasc}}
\newcommand{\gal}[0]{\citetalias{sharma++2019-k2-hermes-age-metallicity-thick-disc}}

\makeatletter
\newcommand\thefontsize[1]{{#1 The current font size is: \f@size pt\par}}
\makeatother

\title[Intrinsic scatter of the scaling relations]{Testing the intrinsic scatter of the asteroseismic scaling relations with \kepler{} red giants}

\author[Li et al.]{%
Yaguang Li,$^{1,2}$\thanks{E-mail: yali4742@uni.sydney.edu.au}
Timothy R. Bedding,$^{1,2}$ 
Dennis Stello,$^{3,1,2}$
Sanjib Sharma,$^{1}$
\newauthor
Daniel Huber$^{4}$ and
Simon J. Murphy$^{1,2}$
\\
$^{1}$Sydney Institute for Astronomy (SIfA), School of Physics, University of Sydney, NSW 2006, Australia\\
$^{2}$Stellar Astrophysics Centre, Department of Physics and Astronomy, Aarhus University, Ny Munkegade 120, \\
DK-8000 Aarhus C, Denmark\\
$^{3}$School of Physics, University of New South Wales, 2052, Australia\\
$^{4}$Institute for Astronomy, University of Hawai`i, 2680 Woodlawn Drive, Honolulu, HI 96822, USA
}
\date{Accepted XXX. Received YYY; in original form ZZZ}

\pubyear{2015}

\begin{document}
\label{firstpage}
\pagerange{\pageref{firstpage}--\pageref{lastpage}}
\maketitle

\begin{abstract}
Asteroseismic scaling relations are often used to derive stellar masses and 
radii, particulaly for stellar, exoplanet, and Galactic studies. It is therefore important that their precisions are known.  
Here we measure the intrinsic scatter of the underlying seismic scaling relations 
for \Dnu{} and \numax{}, using two sharp features that are formed in the 
H--R diagram (or related diagrams) by the red giant populations. 
These features are the edge near the zero-age core-helium-burning phase, 
and the strong clustering of stars at the so-called red giant branch bump. 
The broadening of those features is determined by factors including 
the intrinsic scatter of the scaling relations themselves, 
and therefore it is capable of imposing constraints on them.
We modelled \kepler{} stars with a \galaxia{} synthetic population, 
upon which we applied the intrinsic scatter of the scaling relations to 
match the degree of sharpness seen in the observation.
We found that the random errors from measuring \Dnu{} and \numax{} provide the
dominating scatter that blurs the features. 
As a consequence, we conclude that the scaling relations have 
intrinsic scatter of 
$\sim0.5\%$ (\Dnu{}), $\sim1.1\%$ (\numax{}), $\sim1.7\%$ ($M$) and $\sim0.4\%$ ($R$),
for the SYD pipeline measured \Dnu{} and \numax{}.
This confirms that the scaling relations are very powerful tools. 
In addition, we show that standard evolution models fail to 
predict some of the structures in the observed population of both the HeB and RGB stars. 
Further stellar model improvements are needed to reproduce the exact distributions.

\end{abstract}





\section{Introduction}



The asteroseismic scaling relations for red giants have so far proved to be an extremely useful tool to obtain stellar masses and radii. A critical issue associated with the scaling relations is that their limits are poorly understood \citep{hekker-2020-sc-review}. The intrinsic scatter of the scaling relations, originating from potential hidden dependencies not accounted for in the current relations, can cause a seemingly random fluctuation. Testing the intrinsic scatter of these relations is the aim of this paper.

The scaling relations rely on two characteristic frequencies in the power spectra of solar-like oscillations. The first one is \Dnu{}, the large separation of p modes, approximately proportional to the square root of mean density \citep{ulrich-1986-age}:
\beq\label{eq:sc-dnu}
\frac{\Delta\nu}{\Delta\nu_{\astrosun}} \approx \left(\frac{M}{M_{\astrosun}}\right)^{1/2} \left(\frac{R}{R_{\astrosun}}\right)^{-3/2}.
\eeq
 The second is \numax{}, which is the frequency where the power of the oscillations is strongest. It relates to the surface properties $g/\sqrt{T_{\rm eff}}$ \citep{brown++1991-dection-procyon-scaling-relation,kjeldsen+1995-scaling-relations}: 
\beq\label{eq:sc-numax}
\frac{\nu_{\rm max}}{\nu_{\rm max,\astrosun}} \approx \left(\frac{M}{M_{\astrosun}}\right) \left(\frac{R}{R_{\astrosun}}\right)^{-2} \left(\frac{T_{\rm eff}}{T_{\rm eff,\astrosun}}\right)^{-1/2}.
\eeq
Using these, the mass and radius can be determined if the effective temperature is known \citep{stello++2008-wire-kgiants,kallinger++2010-rg-corot-mass-radius}:
\beq\label{eq:sc-mass}
\frac{M}{M_{\astrosun}} \approx \left(\frac{\nu_{\rm max}}{\nu_{\rm max,\astrosun}}\right)^3 \left(\frac{\Delta\nu}{\Delta\nu_{\astrosun}}\right)^{-4} \left(\frac{T_{\rm eff}}{T_{\rm eff,\astrosun}}\right)^{3/2},
\eeq
\beq\label{eq:sc-radius}
\frac{R}{R_{\astrosun}} \approx \left(\frac{\nu_{\rm max}}{\nu_{\rm max,\astrosun}}\right) \left(\frac{\Delta\nu}{\Delta\nu_{\astrosun}}\right)^{-2} \left(\frac{T_{\rm eff}}{T_{\rm eff,\astrosun}}\right)^{1/2}.
\eeq

From a theoretical point of view, a more accurate value for \Dnu{} can be calculated from oscillation frequencies given a stellar model; thus it is possible to map the departure of Eq.~\ref{eq:sc-dnu}, as a function of [M/H], $M$, \teff{} and evolutionary state \citep{white++2011-asteroseismic-diagrams-cd-epsilon-deltaP-models,sharma++2016-population-rg-kepler,guggenberger++2016-metallicity-scaling-relation,rodrigues++2017-dpi-modelling,serenelli++2017-apokasc-dwarf-subgiant,pinsonneault++2018-apokasc}. 
Improvements are seen when adopting this revised theoretical $\Delta\nu$ over the standard density scaling \citep[e.g.][]{brogaard++2018-accuracy-scaling-relation}. However, there are some degrees of uncertainty. \cite{jcd++2020-aarhus-rgb-osc} found a 0.2\% spread in the theoretical departure stemming from implementing the calculation with different codes, and the degree of model-dependency on physical processes has not been explored extensively. 

The \numax{} scaling relation is much harder to assess theoretically because calculating \numax{} would require a detailed treatment of non-adiabatic processes, via either 1D or 3D stellar models \citep[e.g.][]{balmforth-1992-pulsation-stability-1-mode-thermodynamics,houdek++1999-amplitudes-solarlike,belkacem++2019-3d-rhd,zhou++2019-3d-amplitude-solar-oscillation}. Some works concluded a possible departure could correlate with, for example, the Mach number \citep{belkacem++2011-physics-under-numax-nuc}, magnetic activity \citep{jimenez++2011-nuac-sun-cycle} and mean molecular weight \citep{jimenez++2015-nuac-six-kepler,yildiz++2016-sc-gamma1,viani++2017-numax-sc-metal}. In general, it is still impossible to accurately predict \numax{} from theory. 

Another way to test the scaling relations is by comparing with fundamental data from independent observations. This requires masses and radii obtained by other means, such as astrometric surveys, where radii are deduced using the Stefan-Boltzmann law, eclipsing binaries, where masses and radii are derived from dynamic modelling.
So far, the radii tests based on parallaxes suggest agreement within 4\% for stars smaller than 30\,\rsolar{} \citep{silvaaguirre++2012-seismic-parallax,huber++2017-seismic-radii-gaia,sahlholdt+2018-gaiadr2-sc-radius-dwarfs,hall++2019-rc-gaiadr2-seismo,khan++2019-gaiadr2-zero-point,zinn++2019-radius-sc}. 
With 16 eclipsing binaries, \citet{gaulme++2016-eb-sc} found the asteroseismic masses and radii are systematically overestimated, by factors of 15\% and 5\%, respectively. This result is in disagreement with Gaia radii, possibly because the binary temperature is affected by blending \citep{huber++2017-seismic-radii-gaia,zinn++2019-radius-sc}.
Subsequent analyses indicate that the main source of departure could come from the \Dnu{} scaling relation \citep{brogaard++2018-accuracy-scaling-relation,sharma++2019-k2-hermes-age-metallicity-thick-disc}.

As we noted earlier, the random departures of the scaling relations can be associated with unaccounted factors, 
for example, metallicity, rotation and magnetism, some of which are known to have a wide-ranging distribution among red giants \citep[e.g.][]{mosser++2012-rg-core-rotation-spin-down,stello-2016-supression-rg-kepler,ceillier++2017-surface-rotation-rg-kepler}. 
They could be responsible for some intrinsic scatter in these rather simple relations.

We propose a new approach to investigate the intrinsic scatter, based on two sharp features in the H--R diagram observed among the red giant population. 
The first feature is the accumulation of stars at the bump of red-giant-branch (RGB). The second feature is the sharp edge formed by the zero-age sequence of core-helium-burning (HeB) stars. 
These features were known before seismic observations became available. 

The RGB bump is an evolutionary stage where a star ascending the RGB temporarily drops in luminosity before again ascending towards the tip of the RGB, causing a hump in the luminosity distribution. This feature is prominent in colour-magnitude diagrams of stellar clusters \citep{iben-1968-m15,king++1985-tuc47}. The luminosity drop takes place after the first dredge-up and is caused by a change in the composition profile near the hydrogen-burning shell, leading to a decrease in mean molecular weight outside the composition discontinuity point \citep{refsdal+1970-shell,jcd-2015-rgb-bump}. \kepler{} data show that this bump is also present in the distributions of \Dnu{} and \numax{} \citep{kallinger++2010-kepler-rg-4months,khan-2018-rgb-bump-constraints-envelope-overshooting}.

After reaching the tip of RGB, stars strongly decrease in luminosity and commence core helium burning, forming the red clump, also commonly recognised as the horizontal branch in metal-poor clusters \citep{cannon-1970-rc-discovery,girardi++2010-ngc419-src}. The low-luminosity edge defines the beginning of the red clump and secondary clump phase, which we we will refer to as the zero-age HeB (ZAHeB) phase. This feature is also imprinted on seismic observables \citep{kallinger++2010-kepler-rg-4months,huber++2010-800-rg-kepler,mosser++2010-rg-seismic-property-corot,yuj++2018-16000-rg}.

The fact that the seismic parameters (\Dnu{} and \numax{}) preserve these sharp features indicates that the seismic parameters must be tightly related to the fundamental stellar parameters. Put another way, if there were a large intrinsic scatter in the scaling relations, the features in the seismic diagrams would not be as sharp. Using this principle, we can quantify the limits on the intrinsic scatter in the scaling relations. That is the aim of this paper.


%
\begin{figure}
\includegraphics[width=\columnwidth]{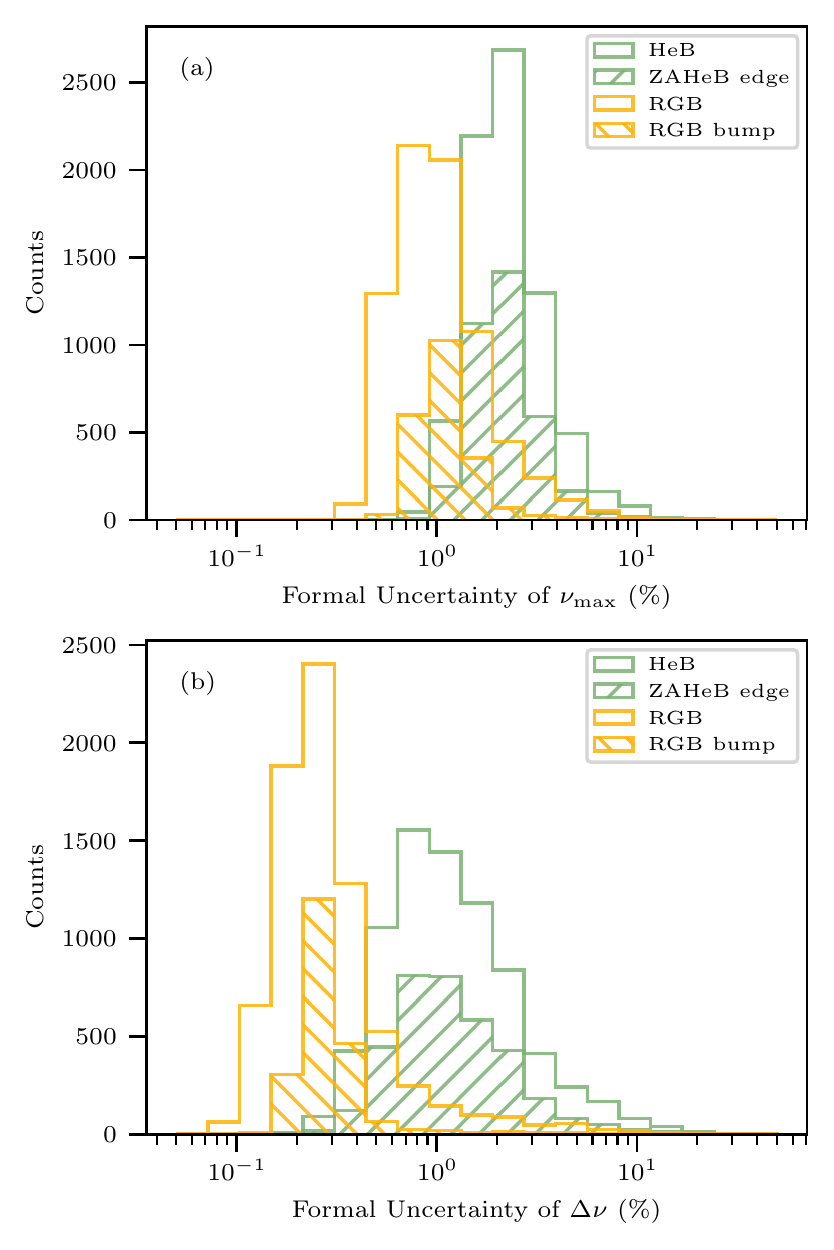}
\caption{Distributions of the \yu{} formal uncertainties of the \numax{} and \Dnu{} measurements.}
\label{fig:formal-error}
\end{figure}

\begin{figure*}
\includegraphics[width=\textwidth]{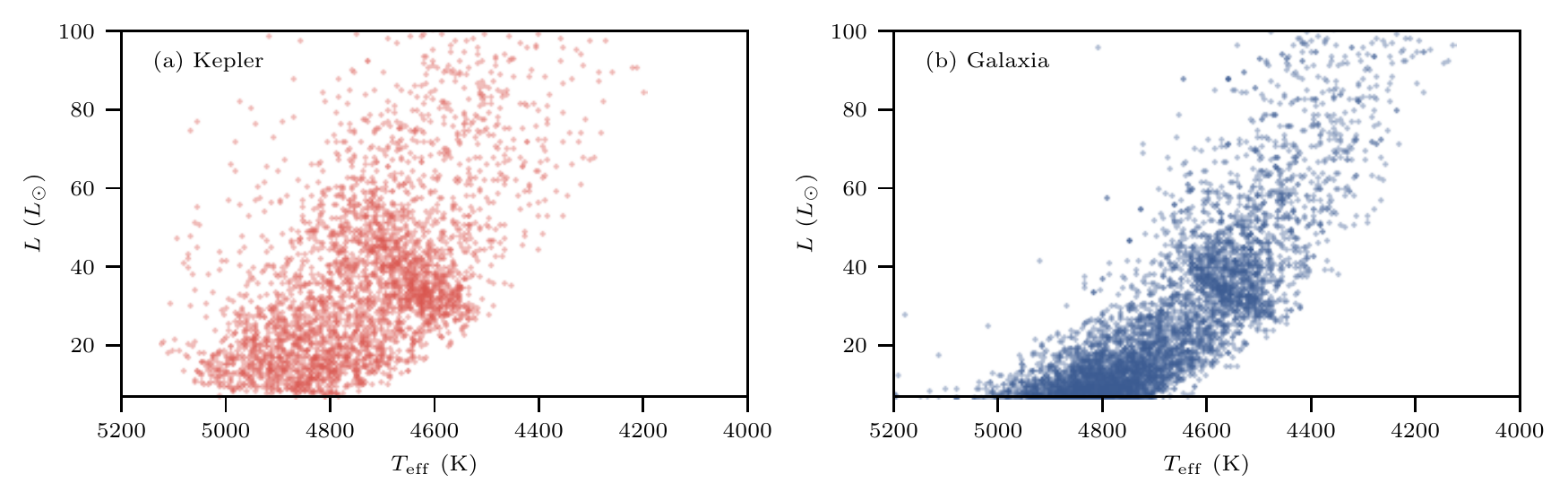}
\includegraphics[width=\textwidth]{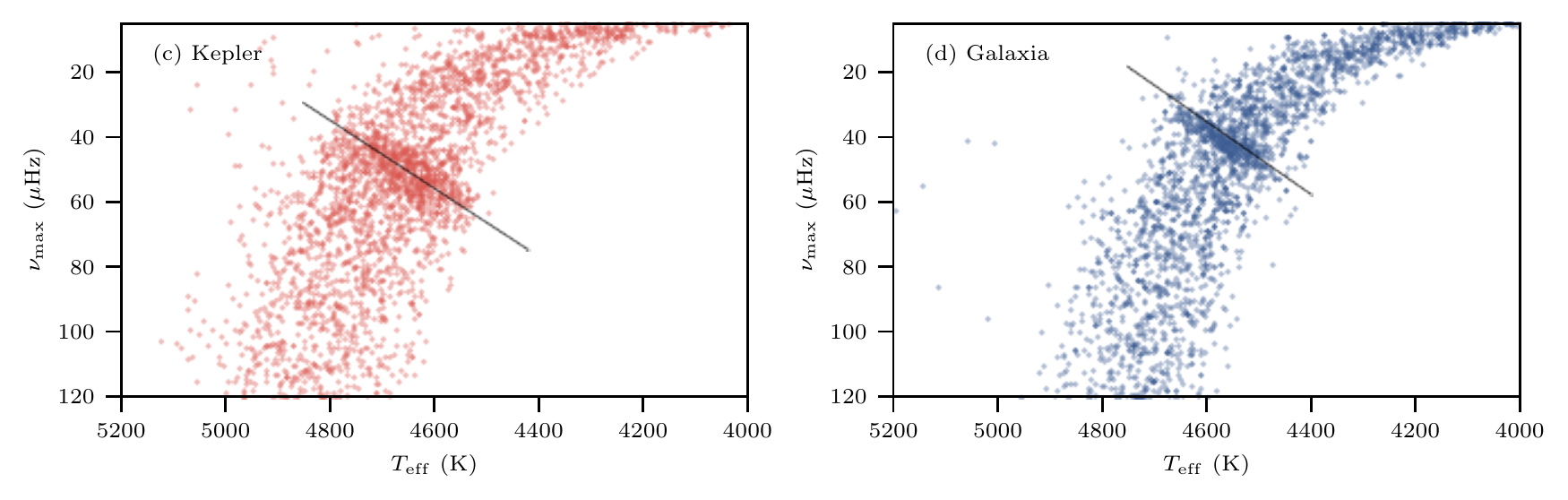}
\includegraphics[width=\textwidth]{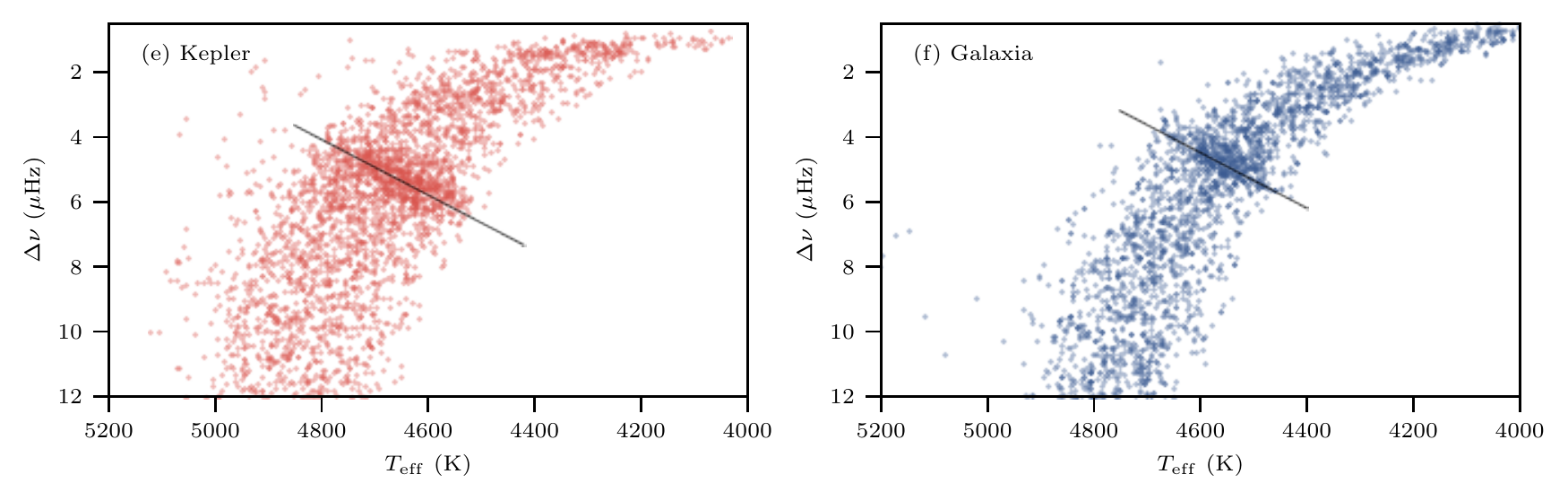}
\includegraphics[width=\textwidth]{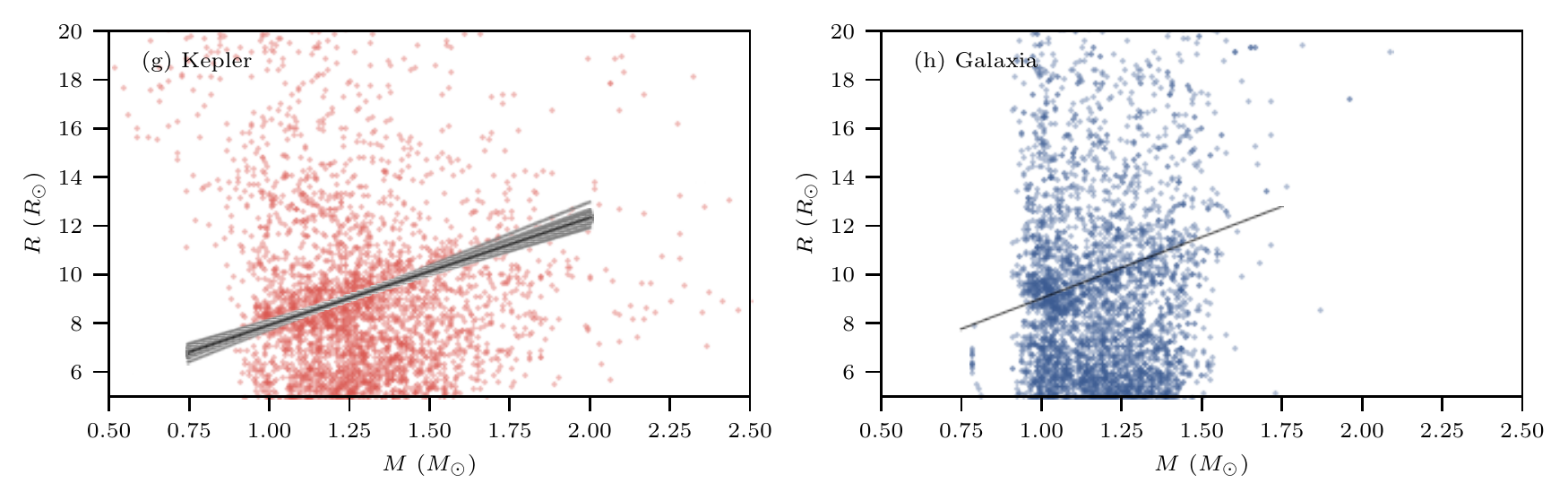}
\caption{$L$ vs. \teff{} (panels a--b), \numax{} vs. \teff{} (panels c--d),
\Dnu{} vs. \teff{} (panels e--f) and $R$ vs. $M$ (bottom g--h) 
for RGB stars in the \apk{} sample (red) and the \gal{} sample (blue). 
The RGB bumps were defined using the black straight fiducial lines. The grey-shaded areas denote the uncertainty of identifying the bump (see \ref{subsubsec:identify-features}).}
\label{fig:diagram-rgb}
\end{figure*}

\begin{figure}
\includegraphics[width=\columnwidth]{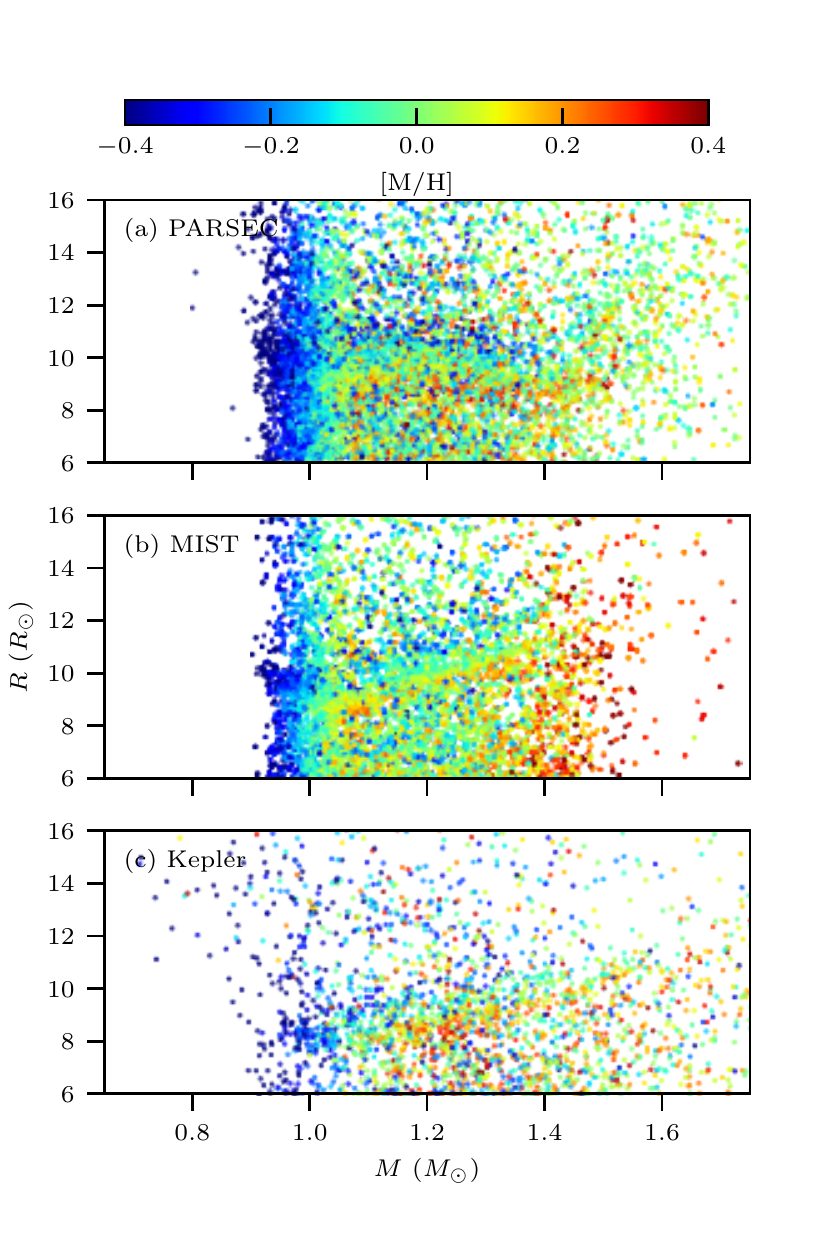}
\caption{Radius vs. mass for RGB stars near the RGB bump, colour-coded by metallicity. The PARSEC and MIST isochrones predict different outcomes on the shape of the RGB bump. }
    \label{fig:rgbb-isochrone}
\end{figure}

\section{Sample selection}
\label{sec:sample}


To create our sample we used the red giants observed by \kepler{}, 
with \Dnu{} and \numax{} measured by the SYD pipeline \citep{huber++2009-syd-pipeline,yuj++2018-16000-rg}, 
and classifications of evolutionary stage (RGB/HeB) from \citet{hon++2017-deep-learning-rc-rgb}. 
We denote this sample as \yu{}, including 7543 HeB stars and 7534 RGB stars. 
A subset of 2531 HeB and 3308 RGB stars with \teff{} and [M/H] from the APOKASC-2 catalog 
\citep{pinsonneault++2018-apokasc} was also used, denoted as \apk{}. 
In Fig.~\ref{fig:formal-error}, we show the distributions of the formal uncertainties 
of \numax{} and \Dnu{} measured by \yu{}.
The \yu{} sample reports a typical formal uncertainty of 
$2.1\%$ on \numax{} and $1.0\%$ on \Dnu{} in HeB stars, 
and $0.95\%$ on \numax{} and $0.3\%$ on \Dnu{} in RGB stars.




To model the observed population, we used a synthetic sample produced by 
\citet{sharma++2019-k2-hermes-age-metallicity-thick-disc} with a Galactic model, 
\galaxia{} \citep{Sharma++2011-galaxia}. 
Compared to a previous synthetic sample in \cite{sharma++2016-population-rg-kepler},
the synthetic sample we used in this work adds a metal-rich thick disc, 
which improves the overall match with the \kepler{} observation 
\citep{sharma++2019-k2-hermes-age-metallicity-thick-disc}. 
Here we denote this sample as \gal{}.
The \gal{} simulated sample is about ten times larger than the \yu{} sample. 
Each star in the simulated sample is associated with an initial mass, an age, 
and a metallicity, sampled from a Galactic distribution function and passed 
through a selection function tied to the \kepler{} mission. 
Other fundamental stellar parameters (e.g. $M$, $R$ and \teff{}) were estimated 
via two different sets of theoretical isochrones: 
PARSEC \citep{marigo++2017-parsec} and 
MIST \citep{Choi++2016-mist-1-solar-scaled-models}.
Both isochrones include some mass loss along the RGB, 
using the \citet{reimers-1975-mass-loss} prescription 
with an efficiency of $\eta_R=0.2$ (PARSEC) and $\eta_R=0.1$ (MIST),
consistent with the asteroseismology of open clusters 
\citep{miglio++2012-mass-loss-ngc6791-ngc6819}.
The seismic parameters, \Dnu{} and \numax{}, were calculated through the scaling 
relations (Eqs.~\ref{eq:sc-dnu} and~\ref{eq:sc-numax}) without any corrections. 
By examining the sharpness of the two features discussed above, and comparing 
the \galaxia{} simulation with the observations, we are able to draw conclusions 
about the intrinsic scatter of the scaling relations.

\begin{figure}
\includegraphics[width=\columnwidth]{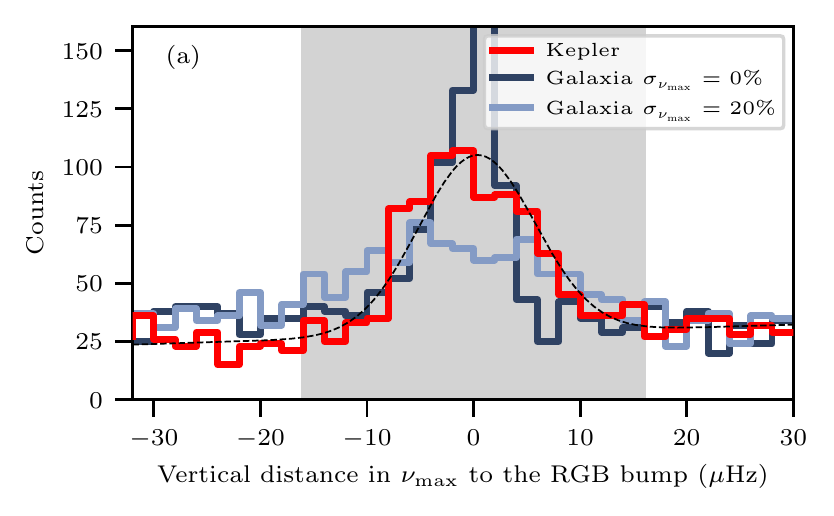}
\includegraphics[width=\columnwidth]{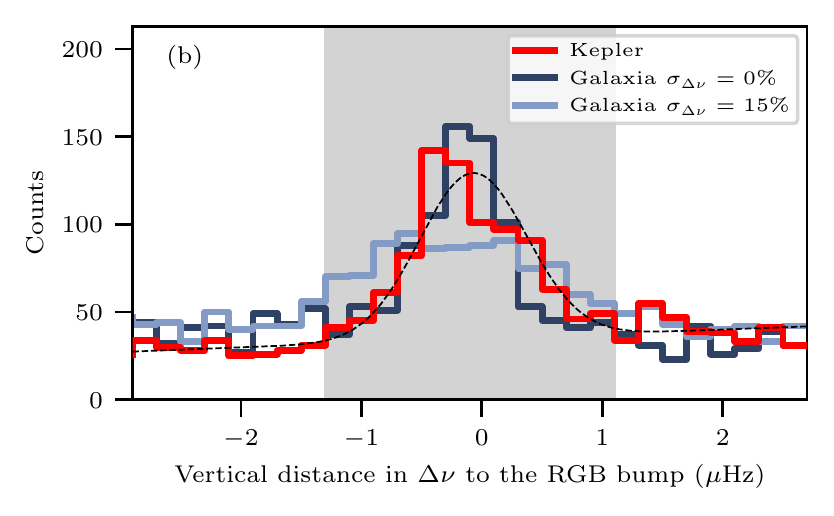}
\includegraphics[width=\columnwidth]{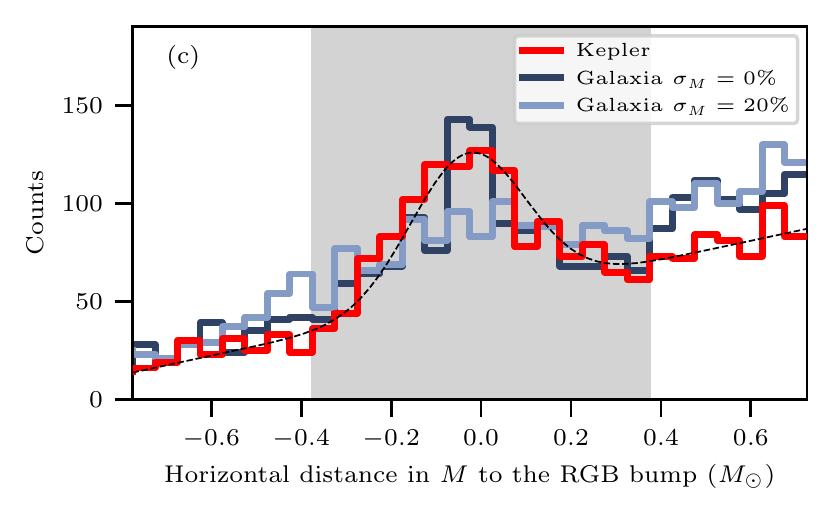}
\includegraphics[width=\columnwidth]{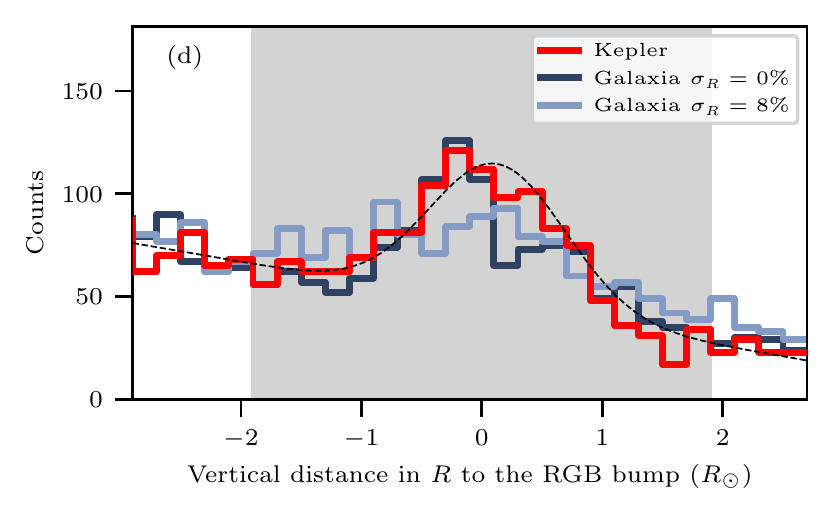}
\caption{Distributions of distances to the bump features. The top two panels are measured in the \teff{}--\numax{} and \teff{}--\Dnu{} diagrams, and the bottom two panels are measured in the $M$--$R$ diagram. The \kepler{} (\apk{}) distributions are shown in red, fitted with a Gaussian model, denoted by the black dashed lines. The synthetic \gal{} samples are shown in blue. The grey-shaded areas denote the range used to compare the data. }
    \label{fig:hist-rgb}
\end{figure}

\section{The RGB bump}
\label{sec:rgbb}

In this section, we look at the RGB bump. 
In a traditional H--R diagram, the bump is tilted so that 
the luminosity $L$ of the bump is a function 
of \teff{} and its shape can be parameterized by stellar mass $M$, 
using $L=L(M)$ and $T_{\rm eff}=T_{\rm eff} (M)$. 
By introducing the \numax{} scaling relation, we can obtain 
$\nu_{\rm max} \propto M L^{-1} T_{\rm eff} ^{7/2}$. 
Therefore, a narrow bump in the $L$--\teff{} plane 
will also show a bump due to this $M$ dependence in the \numax{}--\teff{} plane. 
If the \numax{} scaling relation has some intrinsic scatter due to 
other dependencies, such as metallicity, then the observed bump in 
the \numax{}--\teff{} plane could be wider. 

For \Dnu{}, the argument is similar. 
Fig.~\ref{fig:diagram-rgb} shows the RGB bump for both \kepler{} and \galaxia{} samples.
Here we wish to model the width of the RGB bump. We will start by investigating the features in the \Dnu{}--\teff{} and \numax{}--\teff{} diagrams, and then in the $M$--$R$ diagram.

We further note that the width of RGB bump 
strongly depends on how the physical processes are modelled in the isochrones. 
This is illustrated in Fig.~\ref{fig:rgbb-isochrone}, 
where the shapes of the RGB bump predicted 
by the two sets of isochrones are inconsistent. 
The PARSEC models predict that the stellar radii at the RGB bump should decrease with masses for masses larger than $\sim$1.2 $M_{\astrosun}$. 
However, the opposite is observed in the \kepler{} samples, 
and this behaviour is correctly described by the MIST models.
It implies that the RGB bump may not serve as a useful diagnostic for the scaling relations. 
We will examine this caveat more extensively in section~\ref{subsubsec:model-dependency}. 
Nevertheless, here we still use the RGB bump to introduce our method and we analyse the \gal{} samples with the two isochrones separately.

\subsection{Modelling method}
\label{subsec:rgbb-method}

We used a forward-modelling approach by constructing synthetic 
samples based on the \gal{} sample, and setting the intrinsic 
scatter of the scaling relations, $\sigma$, as a free parameter. 
The width of the bump was evaluated by measuring the distances 
of model samples to the centre of the bump, and fitting their 
distributions to the \apk{} sample. 


The first step was to define the locations of the RGB bump 
in the \apk{} and \gal{} samples
with straight lines in the \numax{}--\teff{} 
and \Dnu{}--\teff{} diagrams, shown in Fig.~\ref{fig:diagram-rgb}. 

We generated a synthetic population by adding random scatter 
to the \gal{} sample. 
Each physical quantity $x$ (one of \Dnu{}, \numax{}, $M$ or $R$) 
for the $i$-th star in the sample was
\beq \label{eq:perturb} x_i = x_{{\textit{Galaxia}}, i} (1 + \sigma_{{\rm total},i}). \eeq %
The quantity $x_{{\textit{Galaxia}},i}$ is the physical value without any perturbation. 
For $M$ and $R$, they were directly estimated from isochrones. 
Note that $M$ is the actual mass rather than the initial mass.
Values for \Dnu{} and \numax{} were determined via scaling relations 
(Eqs.~\ref{eq:sc-dnu} and~\ref{eq:sc-numax}) and further corrected using oscillation 
frequencies (\Dnu{} in particular, see section~\ref{subsec:rgbb-seismo}). 
We modelled the total scatter needed to reproduce the width of the RGB bump, 
$\sigma_{{\rm total},i}$, which was drawn from a normal distribution with 
a standard deviation $\sigma_{\rm total}$.

To account for the scatter induced by the formal uncertainties of the \Dnu{} and \numax{} measurements, 
we modelled each quantity $x$ with
\beq \label{eq:perturb-sc} x_i = x_{{\textit{Galaxia}},i} (1 + \sigma_{x,i} +  \sigma_{{\rm SR},i}), \eeq %
where $\sigma_{x,i}$ represents the fractional uncertainty of $x_{{\textit{Galaxia}},i}$, 
and was drawn randomly from the \apk{} formal uncertainty distribution of RGB bump stars. 
The intrinsic scatter in the scaling relation was modelled via $\sigma_{{\rm SR},i}$, 
drawn from a normal distribution with a standard deviation $\sigma_{{\rm SR}}$. 

We then calculated the distributions of distances to bump lines. 
The bump lines, shown in Fig.~\ref{fig:diagram-rgb}, were picked 
so that the distances to the line have the smallest standard deviation.
For \Dnu{} and \numax{}, we calculated the vertical distances 
in the \Dnu{}--\teff{} and \numax{}--\teff{} diagrams, respectively. 
For $M$ and $R$, we used the horizontal and vertical distances in the $M$--$R$ diagram. 
This procedure allowed us to investigate the scatter in each relation 
separately, because perturbing the horizontal value will not change the 
vertical value, and vice versa. 
In Fig.~\ref{fig:hist-rgb}, we plot the distributions of those distances 
with two representative choices for $\sigma_{\rm{total}}$. 
A larger value for $\sigma_{\rm{total}}$ flattens the hump, demonstrating the width 
of the bump itself provides a measure of the intrinsic scatter in the scaling relations.

Next, we introduce our fitting strategy to enable the comparison, 
which is to match the counts in each bin of the histograms. 
We first identified a central region in the histograms of the 
\apk{} sample by fitting a Gaussian profile plus a sloping straight line, illustrated 
by the dashed curves in Fig.~\ref{fig:hist-rgb}. 
The central region was defined to be a range centred around the Gaussian, 
with a width of 6 times the Gaussian standard deviation. 
In Fig.~\ref{fig:hist-rgb}, they are shown in grey-shaded areas. 
In our fit described below, we matched the distributions in the central regions only.

Because the \gal{} sample is larger than the \apk{} sample, 
we re-scaled the number of model samples by normalising according to 
the \apk{} sample in the central region. 
We also added a constant $c$ as a free parameter to the distance of 
the model samples, in order to compensate for a possible offset of maxima,
which could originate from a bias in identifying the bump. 

We optimised the likelihood function, assuming the distribution of 
counts in each bin is set by Poisson statistics:
\beq \ln L  = \sum_{m_j\neq 0}\left[ d_j \ln m_j - m_j - \ln(d_j!) \right], \eeq
where $d_j$ and $m_j$ are counts in the $j$-th bins of the \kepler{} and model distributions. 
This fitting method is commonly used in population studies to constrain 
the star formation history, initial mass function and binary properties 
\citep[e.g.][]{dolphin-2002-smh-method,geha++2013-imf-dwarf-galaxy,el-badry-2019-twin-binary}.
The posterior distributions of parameters $c$ and $\sigma$ were sampled with 
uninformative flat priors, using a Markov Chain Monte Carlo (MCMC) method.
We used 200 white walkers, burned-in for 500 steps to reach convergence and 
then iterated for another 1000 steps. 
The medians and 68\% credible uncertainties of the parameters were estimated from 
the posteriors directly.

\begin{figure}
\includegraphics[width=\columnwidth]{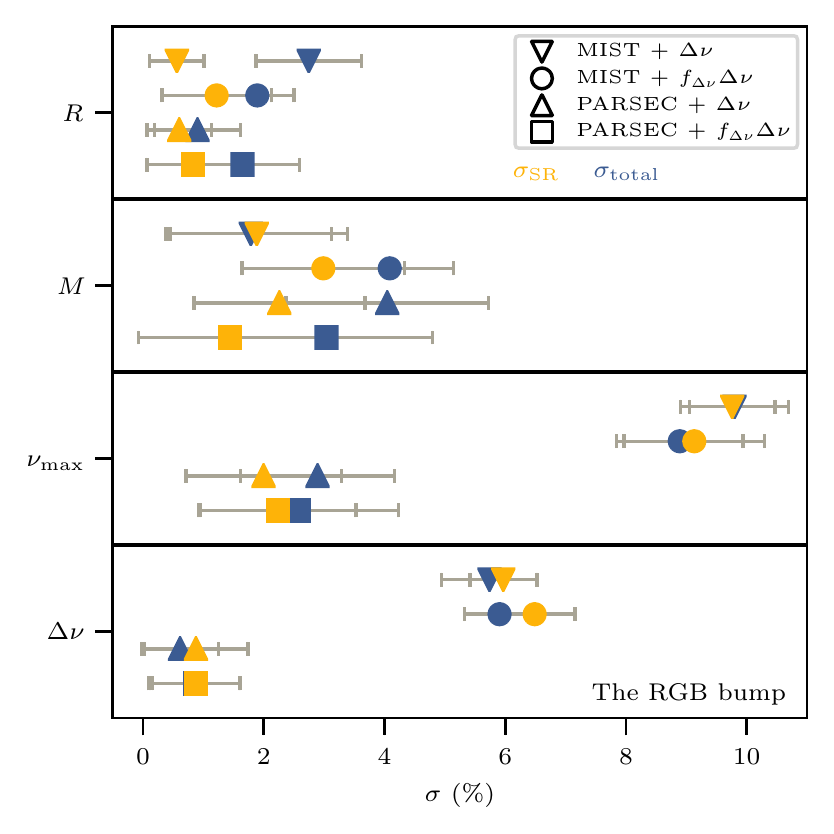}
\caption{Intrinsic scatter of the scaling relations $\sigma_{\rm SR}$ (yellow) and total scatter $\sigma_{\rm total}$ (blue), derived using the width of the RGB bump.}
\label{fig:limits-rgb}
\end{figure}

\begin{figure*} 
\includegraphics[width=\textwidth]{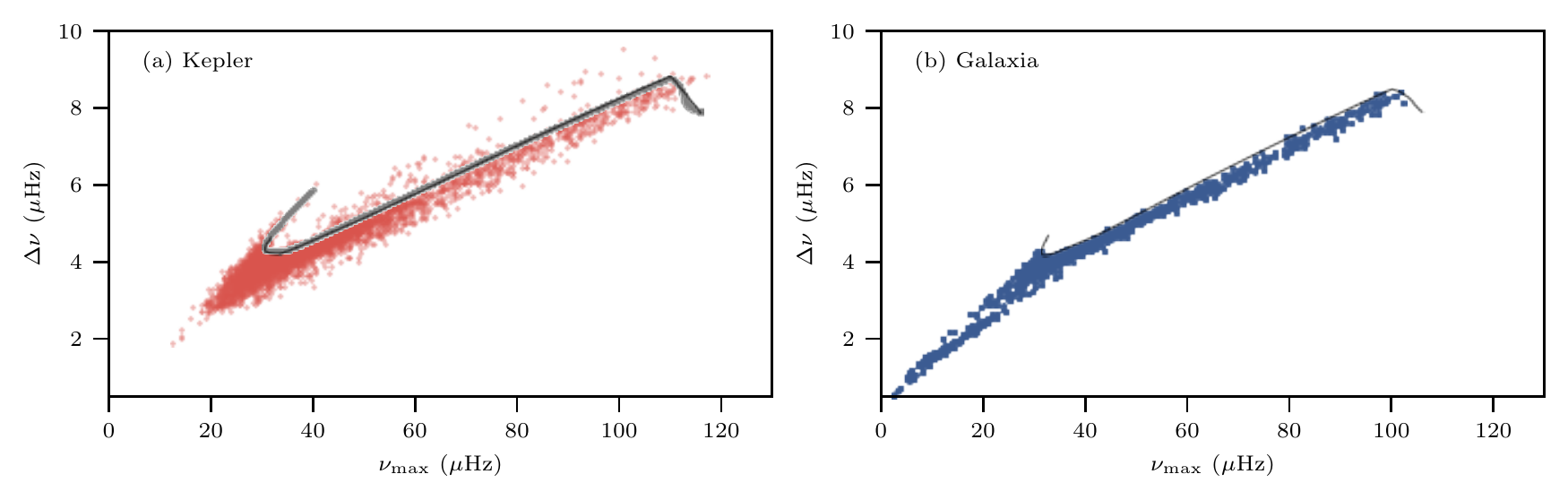}
\includegraphics[width=\textwidth]{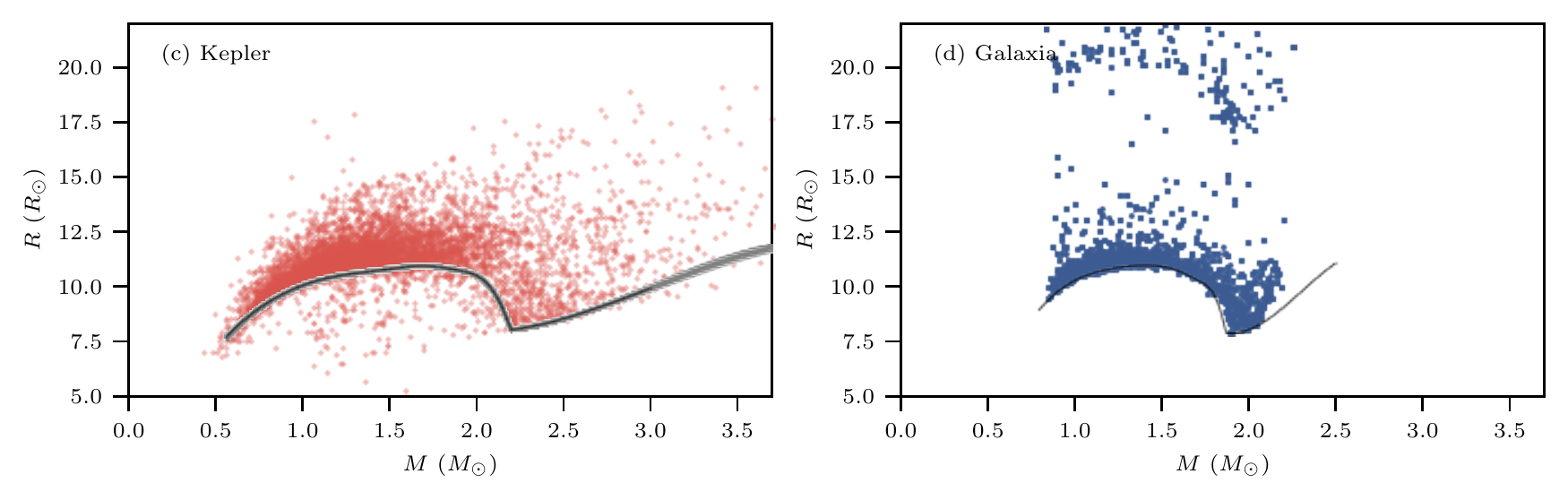}
\caption{\numax{}--\Dnu{} diagram (panels a--b) and $M$--$R$ diagram (panels c--d) 
for HeB stars in the \yu{} sample (red) and the \gal{} sample (blue). 
The zero-age HeB edges were defined using splines, shown as the black lines. 
The grey-shaded areas denote the uncertainty of identifying the edges (see section~\ref{subsubsec:identify-features}).
The stars around $20$ $R_{\astrosun}$ in the \gal{} sample are at the asymptotic-giant-branch phase.
}
\label{fig:diagram-heb}
\end{figure*}

\subsection{Results}
\label{subsec:rgbb-seismo}
Our first step is to derive the total scatter responsible for the width of the RGB bump, 
$\sigma_{\rm total}$, in Eq.~\ref{eq:perturb}. 
We obtained
$0.61\%$ (\Dnu{}),
$2.89\%$ (\numax{}),
$4.05\%$ ($M$) and
$0.90\%$ ($R$) with PARSEC, and
$5.74\%\pm0.80\%$ (\Dnu{}),
$9.80\%\pm0.90\%$ (\numax{}),
$1.79\%\pm1.34\%$ ($M$) and
$2.75\%\pm0.88\%$ ($R$) with MIST.

Next we took the formal uncertainties of the \Dnu{} and \numax{} measurements into account
and obtained the limits on the intrinsic scatter 
of the scaling relations, $\sigma_{\rm SR}$, in Eq.~\ref{eq:perturb-sc}. 
With PARSEC, we obtained 
$0.88\%$ (\Dnu{}),
$2.00\%$ (\numax{}),
$2.26\%$ ($M$) and
$0.60\%$ ($R$).
With MIST, we obtained
$5.97\%$ (\Dnu{}),
$9.76\%$ (\numax{}),
$1.89\%$ ($M$) and
$0.56\%$ ($R$). 
These numbers are plotted in Fig.~\ref{fig:limits-rgb}. There is a huge difference between MIST and PARSEC. We will discuss it in section~\ref{subsubsec:model-dependency}.

\section{The ZAHeB edge}
\label{sec:zaheb}

Similar to the RGB bump, the zero-age sequence of HeB stars (ZAHeB) also forms a well-defined feature in the H--R diagram \citep{girardi++2010-ngc419-src,girardi-2016-rc-review}. 
We note that the transition from the red clump (low-mass stars that ignite helium in a fully degenerate core) 
to the secondary red clump (higher-mass stars that ignite helium in a partly or non-degenerate core) 
is smooth and continuous \citep{girardi-2016-rc-review}. 
Given the scaling relations, there should exist a close correlation between $\Delta\nu$ and $\nu_{\rm max}$ for the ZAHeB.  

In Fig.~\ref{fig:diagram-heb}, we show the HeB stars in the \Dnu{}--\numax{} and $M$--$R$ diagrams.
The ZAHeB appears as a very sharp feature: all HeB stars are located at only one side of the ZAHeB, forming a remarkably sharp edge. 
\footnote{It has not escaped our attention that Fig~\ref{fig:diagram-heb}(a) bears 
a strong resemblance to the logo of a major footwear manufacturer. 
We plan to investigate sponsorship opportunities.}
Now we use the sharpness of this edge to quantify the intrinsic scatter of the scaling relations.


\subsection{Modelling method}
\label{subsec:heb-method}

\begin{figure}
\includegraphics[width=\columnwidth]{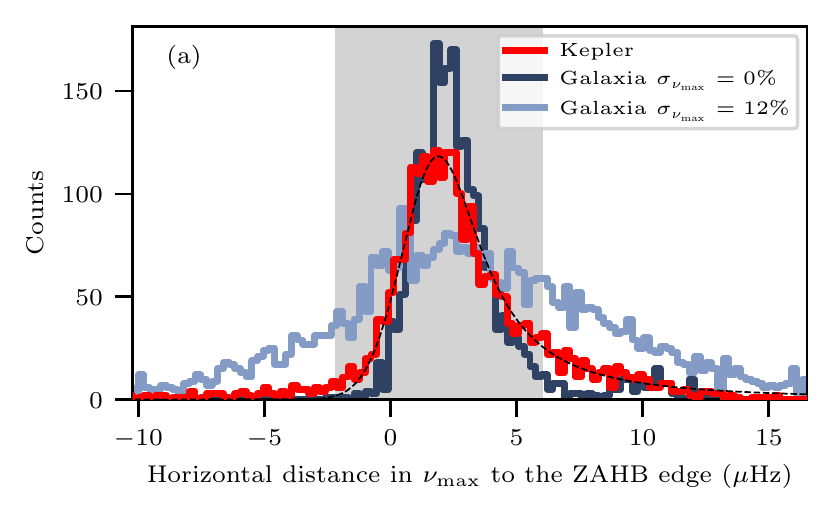}
\includegraphics[width=\columnwidth]{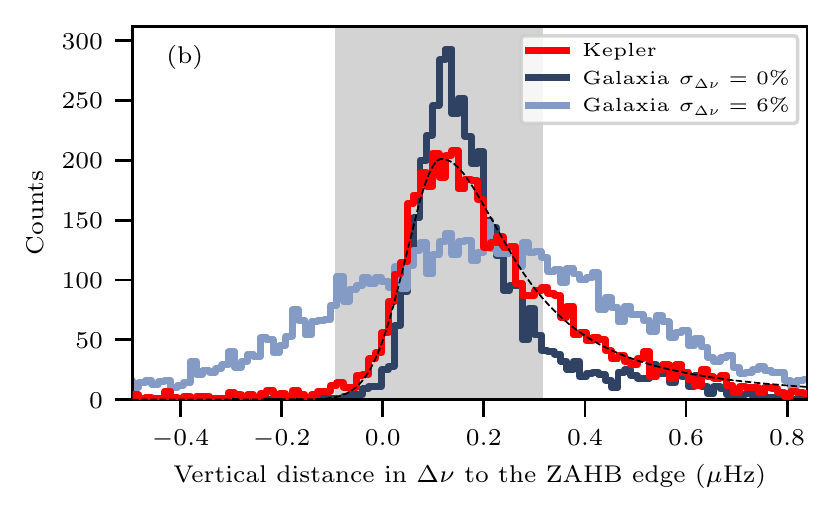}
\includegraphics[width=\columnwidth]{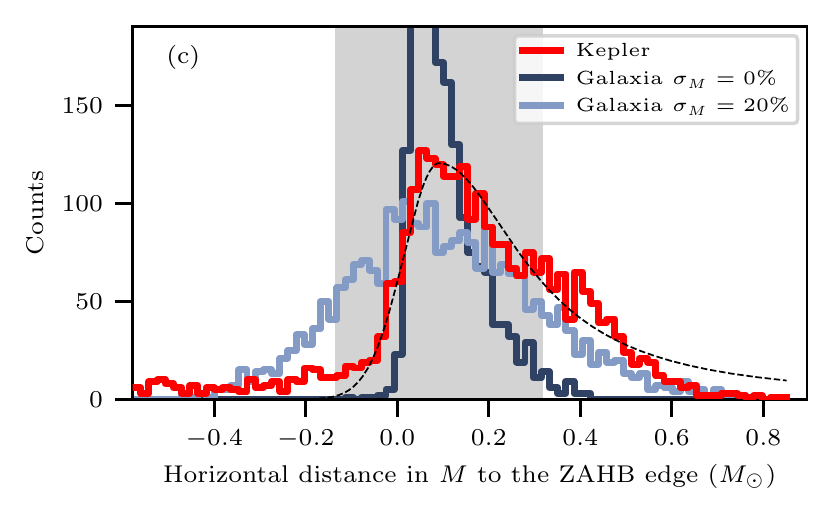}
\includegraphics[width=\columnwidth]{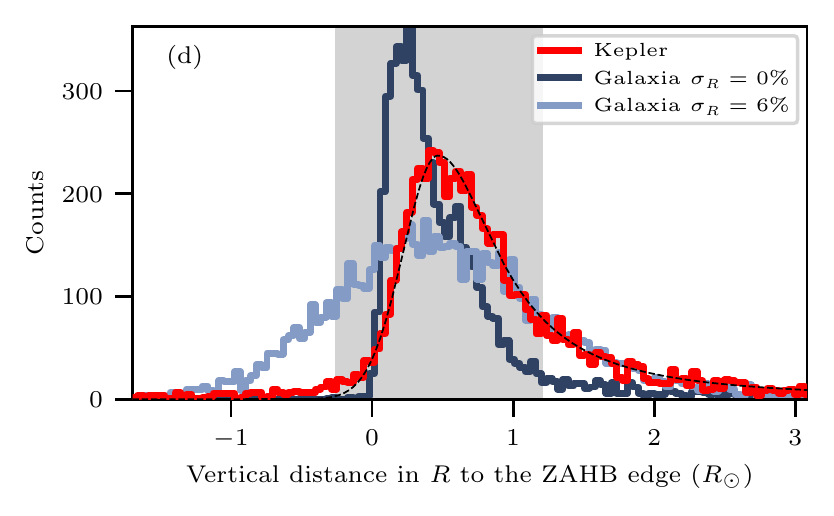}
\caption{Distributions of distances to the ZAHeB edges. The top two panels are measured in the \numax{}--\Dnu{} diagram, and the bottom two panels are measured in the $M$--$R$ diagram. The \kepler{} (\yu{}) distributions are shown in red, fitted with a half-Gaussian-half-Lorentzian model, denoted by the black dashed lines. The synthetic \gal{} samples are shown in blue. The grey-shaded areas denote the range used to compare the data. }
    \label{fig:hist-heb}
\end{figure}

To measure the sharpness of the ZAHeB edge, we used a modelling method similar 
to that for the RGB bump in section~\ref{subsec:rgbb-method}, but with three 
important differences.

The first difference is related to defining the location of the feature.
For RGB stars, we used straight lines to denote the location of the bump.
For HeB stars, we used splines in the \Dnu{}--\numax{} and $M$--$R$ diagrams 
to define the ZAHeB edges. 
This is illustrated in Fig.~\ref{fig:diagram-heb}, where the edges are shown 
as black lines.

The second difference is how we calculated the horizontal and vertical 
distances to the ZAHeB edge.
In the \Dnu{}--\numax{} diagram, the stars below the lowest point of the edge 
do not have a meaningful horizontal distance. We therefore excluded them for 
the horizontal distance calculation. 
The same strategy was also applied to all stars that lie on the left of 
the leftmost point of the defined ZAHeB edge when calculating vertical distances.
Similarly, in the $M$--$R$ diagram, the stars above the highest point of 
the ZAHeB edge were not considered in calculating horizontal distances.
In Fig.~\ref{fig:hist-heb}, we plot the distributions of those distances with 
two $\sigma_{\rm total}$. 
As for the RGB bump, we see that a larger scatter $\sigma_{\rm total}$ smooths the hump.

The third difference is that, in order to choose regions near each edge to compare, 
we fitted a profile to the distributions for the \yu{} sample. 
The profiles, shown as the dashed lines in Fig.~\ref{fig:hist-heb},
consisted of a half Gaussian (left) and a half Lorentzian (right).
The histogram region that we fitted was a range centred at the Gaussian's centre, 
with a width of 6 times the Gaussian's standard deviation. These regions are 
shown as grey-shaded areas.

\subsection{Results}
\label{subsec:heb-seismo}

\begin{figure}
	\includegraphics[width=\columnwidth]{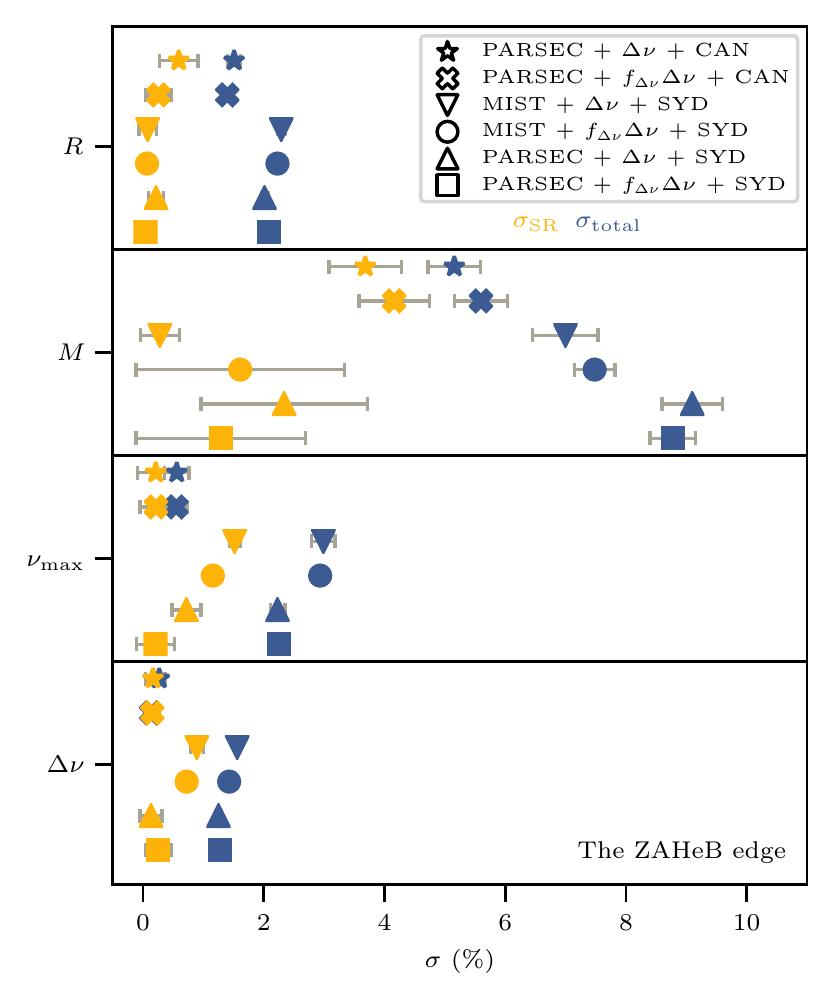}
    \caption{Intrinsic scatter of the scaling relations $\sigma_{\rm SR}$ (yellow) and total scatter $\sigma_{\rm total}$ (blue), derived using the sharpness of the ZAHeB edge.}
    \label{fig:limits-heb}
\end{figure}

We measured the total scatter $\sigma_{\rm total}$ that contributes to the 
broadening of the edges in the \numax{}--\Dnu{} and $M$--$R$ diagrams:
$1.25\%\pm0.05\%$ (\Dnu{}),
$2.23\%\pm0.12\%$ (\numax{}),
$9.10\%\pm0.50\%$ ($M$) and
$2.01\%\pm0.05\%$ ($R$) using the PARSEC models, and
$1.56\%\pm0.04\%$ (\Dnu{}),
$2.99\%\pm0.19\%$ (\numax{}),
$7.00\%\pm0.54\%$ ($M$) and
$2.29\%\pm0.07\%$ ($R$) using the MIST models.
These numbers are in general agreement with the formal uncertainties of \Dnu{} and \numax{}
reported by \yu{} for HeB stars (Fig.~\ref{fig:formal-error}), suggesting a main contribution to 
the broadening of the ZAHeB edge. 

Next, we tested whether we needed to add intrinsic scaling relation scatter to 
the \yu{} measurement uncertainties in order to reproduce the sharpness of 
the ZAHeB edge. We derived $\sigma_{\rm SR}$ with the PARSEC models: 
$0.13\%\pm0.18\%$ (\Dnu{}),
$0.72\%\pm0.24\%$ (\numax{}),
$2.34\%\pm1.38\%$ ($M$) and
$0.22\%\pm0.12\%$ ($R$).
And with the MIST models we obtained
$0.89\%\pm0.11\%$ (\Dnu{}),
$1.52\%\pm0.09\%$ (\numax{}),
$0.28\%\pm0.32\%$ ($M$) and
$0.08\%\pm0.14\%$ ($R$).
These numbers are plotted in Fig.~\ref{fig:limits-heb}. 




\section{Discussion}
\label{sec:disc}

\subsection{Assessing uncertainties}
\label{subsec:uncertainties}

\subsubsection{The uncertainty of modelling the stellar population}
\label{subsubsec:model-dependency}


Figs.~\ref{fig:limits-rgb} and~\ref{fig:limits-heb} present the total scatter $\sigma_{\rm total}$ and 
the limits on the intrinsic scatter of the scaling relations 
$\sigma_{\rm SR}$ derived under various assumptions.
A feature become immediately obvious: the results depend on how the synthetic stars are modelled.

We first discuss its impact on the RGB bump.
The input physics has significantly influenced the width of the RGB bump.  
As we already illustrated in Fig.~\ref{fig:rgbb-isochrone}, 
the shapes of the RGB bump predicted by the two isochrones are inconsistent. 
Furthermore, the PARSEC models predict a wider bump than the observation, 
even when the quantities were not perturbed with any scatter. 
In contrast, the MIST models present a much narrower bump,
and so a much larger scatter needs to be added to match the observed width.

For some cases in Fig.~\ref{fig:limits-rgb}, 
$\sigma_{\rm total}$ exceeds $\sigma_{\rm SR}$,
which is also a signature that the shape of RGB bump predicted by 
models cannot properly match the observation. 
For example, the \gal{} synthetic samples overestimate the number 
of low-mass stars near $\sim1$ $M_{\astrosun}$, which can be seen from 
the panel h of Fig.~\ref{fig:diagram-rgb}. 
The mismatch was first discussed by \citet{sharma++2016-population-rg-kepler}, 
and \citet{sharma++2019-k2-hermes-age-metallicity-thick-disc} used 
a metal-rich thick disc to ease the tension, but the inconsistency still exists. 

The RGB bump is an important diagnostic for stellar physics.
\citet{jcd-2015-rgb-bump} linked the width of the RGB bump with
the magnitude of the hydrogen abundance discontinuity in the vicinity of the 
hydrogen-burning shell, which depends on the evolution history. 
The modelling of convection (e.g. mixing-length and overshoot) can also
have an impact on the location of the bump
(see \citealt{khan-2018-rgb-bump-constraints-envelope-overshooting} and references therein).

From the above discussion, we conclude that the RGB bump is not useful for our purpose,
unless an initial calibration of stellar models is properly done. 
The calibration can be achieved 
by matching luminosity distributions using benchmark data, 
and then the feature can be compared in the seismic diagrams. 
This is beyond the scope of this paper, and we defer it for future work.

Turning into the ZAHeB edge, we noticed the shape of the edges is also model-dependent.
A noticeable feature in Fig.~\ref{fig:diagram-heb} is that the mass limit of the 
helium flash in models does not match with the observation.
The mass limit has been shown to be
dependent on the treatment of overshooting \citep{girardi-2016-rc-review},
which is often considered as a free parameter in stellar modelling. 
Fig.~\ref{fig:diagram-heb} also shows a lack of low-mass HeB stars in the \gal{} sample, 
likely because the synthetic sample
does not incorporate enough mass loss. 

Despite these model uncertainties, we found they are less sensitive to 
the values of $\sigma_{\rm SR}$ that we are interested in.
This means that using the ZAHeB edge 
to put a limit on the intrinsic scatter of the scaling relations 
is a realistic approach in the current work.


\subsubsection{The uncertainty of identifying the features}
\label{subsubsec:identify-features}

The chosen ZAHeB edges and RGB bumps in the \kepler{} samples
might deviate from their real positions. 
Here we test its influence on the inferred $\sigma_{\rm SR}$ by
shifting the locations 
in the observation samples. 
We perturbed the points 
used to define the splines (ZAHeB edge) and the straight lines (RGB bump) 
with an amount of $s/\sqrt{N}$. 
We took $s$ as the standard deviation of the Gaussian profiles fitted in
Figs.~\ref{fig:hist-rgb} and~\ref{fig:hist-heb}, and $N$ as the number of samples. 
This perturbation is similar to the standard deviation of the sample mean, 
and should provide a good approximation to the uncertainty of choosing the center of those features. 
In Fig.~\ref{fig:diagram-rgb} and~\ref{fig:diagram-heb}, 
the grey-shaded areas show the amount of uncertainty.
We found the resulting $\sigma_{\rm SR}$ agrees with the reported values within
$0.06\%$ for \Dnu{}, $0.1\%$ for \numax{}, $1.8\%$ for $M$, and $0.7\%$ for $R$.
This result indicates that the uncertainty of identifying the features is much smaller
than $\sigma_{\rm total}$, but is on a similar level of $\sigma_{\rm SR}$.

In addition, we note that there is a selection effect 
(will be shown in section~\ref{subsec:heb-method} and Fig.~\ref{fig:mass}) 
due to excluding HeB stars near the ZAHeB edge
when there were no horizontal or vertical distances. 
For example, the obtained values
for the mass relation are only applicable to stars in the range of 
$0.8$--$1.1$ $M_{\astrosun}$, so the derived numbers for 
$\sigma_{\rm total}$ and $\sigma_{\rm SR}$ are the averages 
for those specific subsamples, making the numbers between 
each relations not directly comparable. 

\subsubsection{The uncertainty of measuring \Dnu{} and \numax{}}
\label{subsubsec:syd-pip}

The limits we obtained for $\sigma_{\rm SR}$ depend on 
how well the values for \Dnu{} and \numax{} are measured. 
Up to now we focused our discussion using the SYD pipeline, 
which measures \Dnu{} and \numax{} from a global fitting of the 
power spectrum \citep{huber++2009-syd-pipeline,yuj++2018-16000-rg}. 
Although the global fitting method is more common,
an alternative approach is to only use the radial mode frequencies 
and avoid the effect from mixed modes. An example is the CAN pipeline 
\citep{kallinger++2010-kepler-rg-4months},
which obtained a more precise measurements on \Dnu{} and \numax{}. 
For stars near the ZAHeB edge, their typical formal uncertainties are 
0.6\% for \numax{}, and 0.3\% for \Dnu{} \citep{pinsonneault++2018-apokasc}. 
Using their reported uncertainties, 
we show in Fig.~\ref{fig:diagram-heb}, 
that the values for $\sigma_{\rm SR}$ in the \Dnu{} and 
\numax{} relations can greatly decrease. 
If the values for \Dnu{} and \numax{} are measured in this way, 
the scaling relations can have much smaller intrinsic scatter in principle.
However, we also found the intrinsic scatter in the $M$ and $R$ scaling relations 
does not decrease accordingly, because the uncertainty of \teff{} still dominates. 
In the rest of this paper, we continue our discussion using the SYD pipeline values.



\subsection{The intrinsic scatter of the scaling relations}
\label{subsec:scatter}

Based on the discussion in section~\ref{subsec:uncertainties}, 
we estimate the final values of the intrinsic scatter of the scaling relations, $\sigma_{\rm SR}$,
by averaging them from both RGB and HeB stars for the $M$ and $R$ relations,
but only HeB stars for the \Dnu{} and \numax{} relations,
because these values tend to show less severe dependencies on isochrones.
We conclude that the intrinsic scatter of the scaling relations have values of 
$\sim0.5\%$ (\Dnu{}), $\sim1.1\%$ (\numax{}), $\sim1.7\%$ ($M$) and $\sim0.4\%$ ($R$),
for the SYD pipeline, 
keeping in mind that the systematic uncertainty of our method is on a similar level.
The values of $\sigma_{\rm SR}$ are small in general, 
suggesting the observational uncertainty typically exceeds the 
intrinsic scatter of the scaling relations even with 4 yr of \kepler{} data for the SYD pipeline.

In our study, we separately located the ZAHeB edges in the \kepler{} and 
\galaxia{} samples. 
This means that any systematic offset in the scaling relations 
(for example, using a different set of solar reference values) 
would not be reflected in $\sigma_{\rm SR}$. 
The intrinsic scatter in the scaling relation can still be small 
compared to any systematic offset in the scaling relations.

\subsection{Correcting the scaling relations with theoretical models}
\label{subsec:heb-fdnu}
\begin{figure}
	\includegraphics[width=\columnwidth]{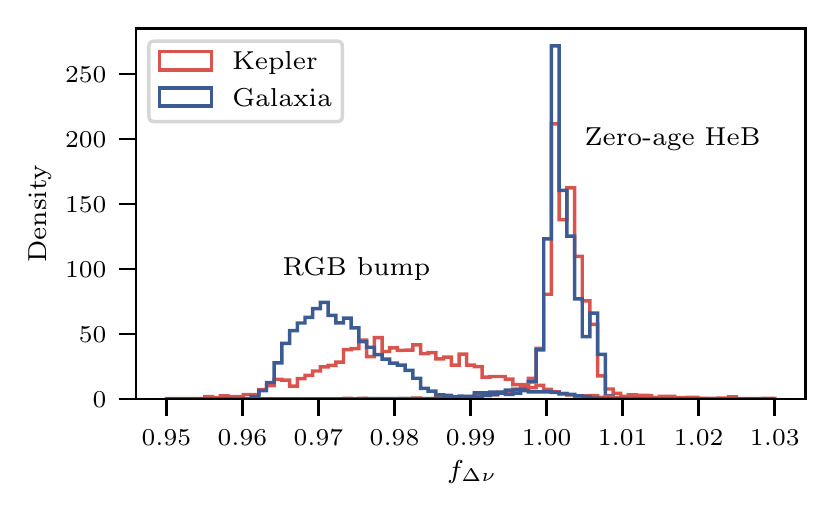}
    \caption{Distributions of the correction factor \fdnu{} for stars near the RGB bump and stars near the ZAHeB edge (grey-shaded area in Fig.~\ref{fig:hist-rgb} and~\ref{fig:hist-heb}) in both \kepler{} (red) and \galaxia{} (blue) samples.}
    \label{fig:fdnu}
\end{figure}

It is interesting to test whether the common model-based 
correction of \Dnu{} proposed by \citet{sharma++2016-population-rg-kepler} 
can reduce the scatter in the scaling relations. 
We calculated the departure of the \Dnu{} scaling relation, \fdnu{}, for 
each star in both samples. 
We implemented the corrected mass $M'=f_{\Delta\nu}^4 M$ and 
radius $R'= f_{\Delta\nu}^2 R$ in the \kepler{} sample, 
and the corrected p-mode separation $\Delta\nu' = f_{\Delta\nu} \Delta\nu$ 
in the synthetic sample. 

These \Dnu{} corrections made little difference to our results, 
for both RGB stars (Fig.~\ref{fig:limits-rgb}) and HeB stars 
(Fig.~\ref{fig:limits-heb}).
The likely explanation is that \fdnu{} mainly corrects the 
systematic offsets in the scaling relations, which affect the 
location of the RGB bump and ZAHeB edge, but has a negligible 
influence on the intrinsic scatter of the scaling 
relations (Fig.~\ref{fig:fdnu}). 
The standard deviation of \fdnu{} for stars near the ZAHeB 
edge is below $0.5\%$, and that for stars near the RGB bump 
is about $1.0\%$. 



\subsection{The intrinsic scatter of the scaling relations as a function of mass and metallicity}
\label{subsec:heb-mass-metal}
\begin{figure}
	\includegraphics[width=\columnwidth]{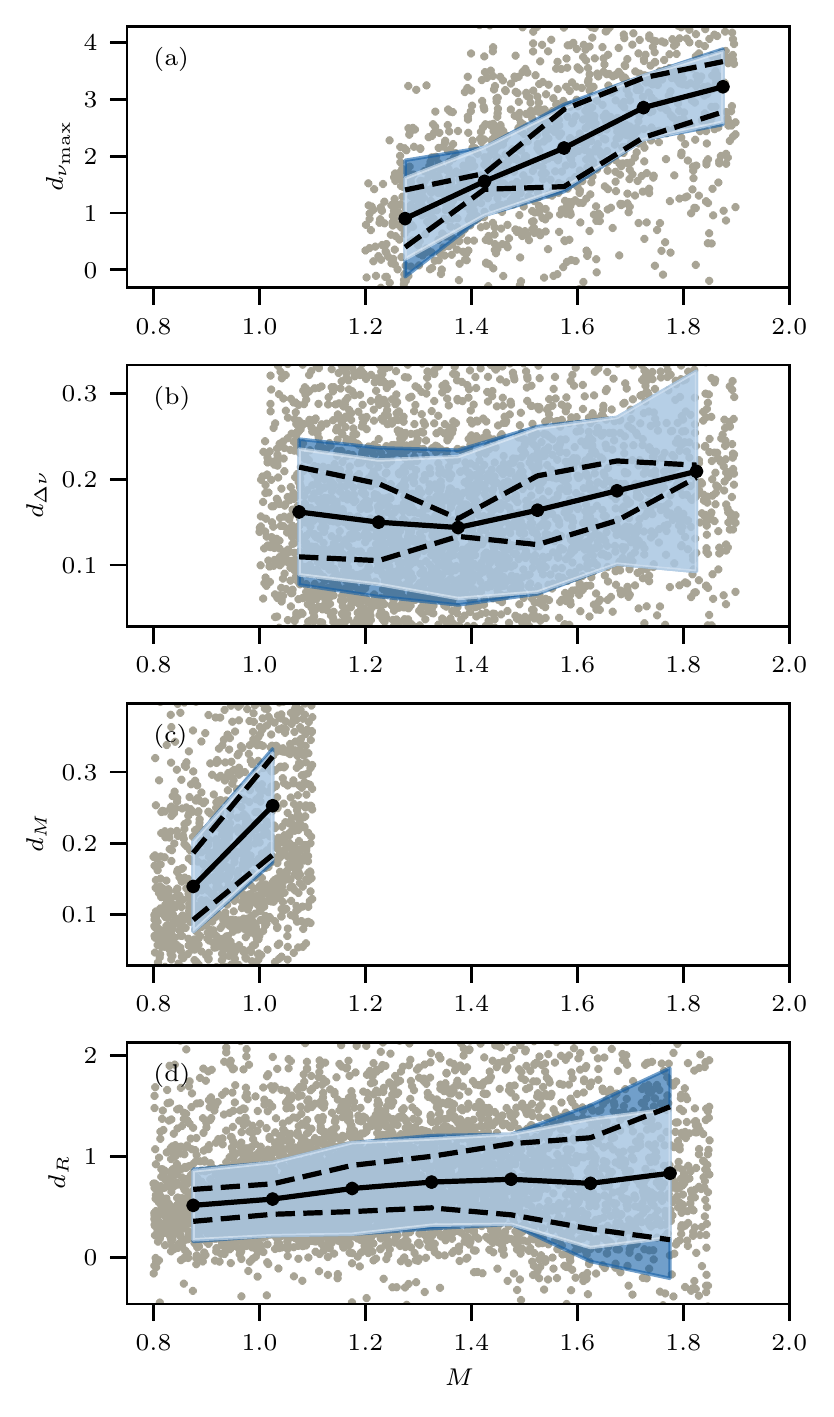}
    \caption{Distances to the ZAHeB edge as a function of stellar mass for the \yu{} sample (grey points). The solid black line traces the median values of the distances in each mass bin. The light blue show the formal uncertainties of \Dnu{} and \numax{} reported by the SYD pipeline, and the dark blue regions show the intrinsic scatter of the scaling relations $\sigma_{\rm SR}$. The dashed black lines show the total scatter $\sigma_{\rm total}$. }
    \label{fig:mass}
\end{figure}

\begin{figure}
	\includegraphics[width=\columnwidth]{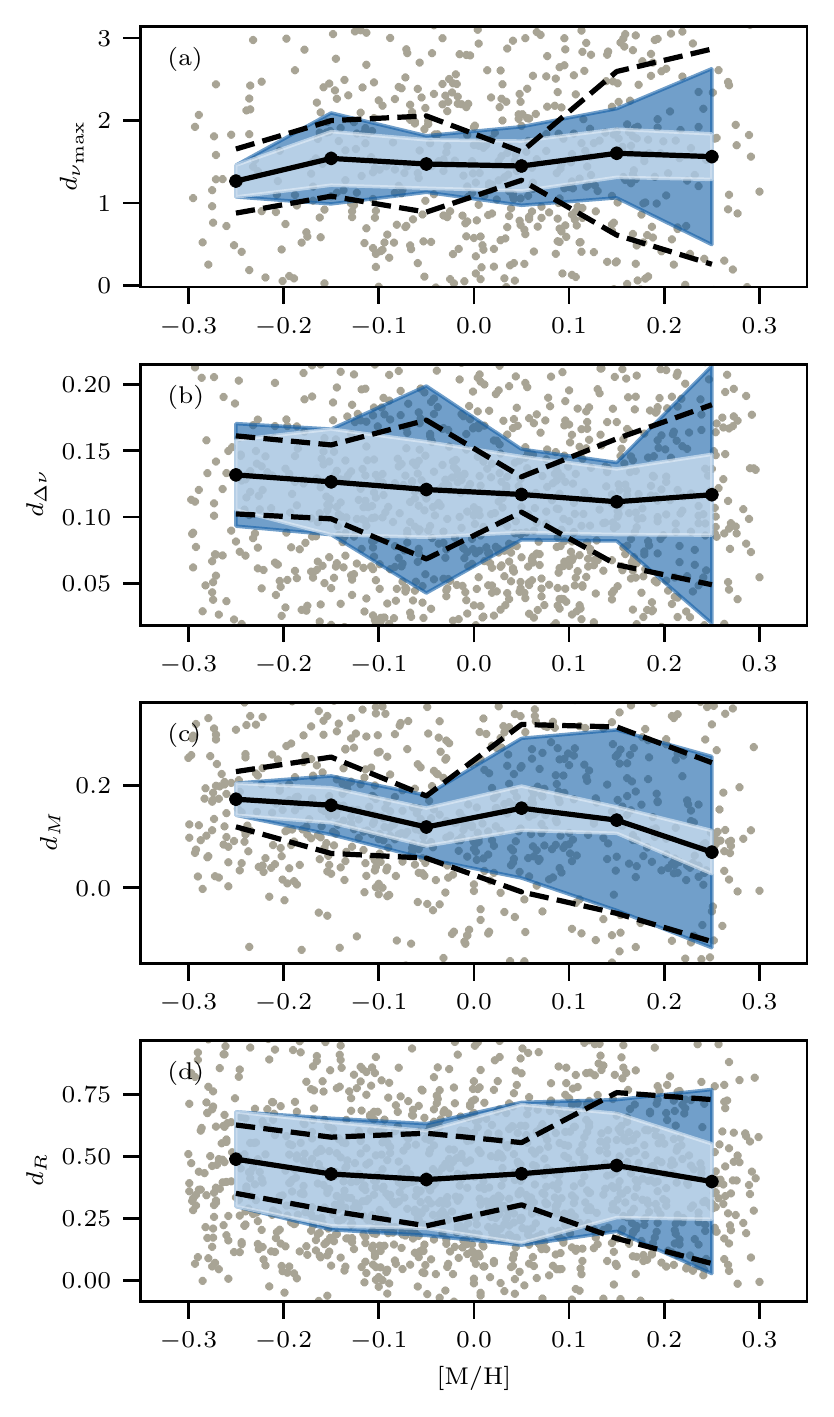}
    \caption{Similar to Fig.~\ref{fig:mass} but shown as a function of metallicity, and restricted to the \apk{} sample.}
    \label{fig:feh}
\end{figure}


We expect the intrinsic scatter in the scaling relations to be a 
function of stellar mass and metallicity, as we see in \fdnu{}. 
To test whether this dependence can be seen in our sample, 
we used HeB stars and divided both the \kepler{} and \galaxia{} samples into bins 
with equal widths in $M$ and [M/H], and repeated the exercise 
in each bin. 
We note that for \Dnu{}, \numax{} and $M$, we could only test 
a limited range in mass, because some points do not have vertical 
or horizontal distances. 
To study the dependence on [M/H], we used the \apk{} 
sample instead of the \yu{} sample, because the \apk{}
metallicities were derived from a single instrument.

In Fig.~\ref{fig:mass} and~\ref{fig:feh} we show 
$\sigma_{\rm total}$ (dark blue regions) and $\sigma_{\rm SR}$ 
(dashed lines) as functions of $M$ and [M/H], respectively. 
We find no obvious change in the spread of points for \Dnu{}, 
\numax{} and $M$, possibly due to a direct consequence of the method uncertainty 
we claimed in section~\ref{subsubsec:identify-features}.
The data also suggest that a higher mass and 
higher metallicity may result in a larger intrinsic scatter for 
the radius scaling relation. 
Whether this is a true statement can be found by populating more stars 
in the high mass and high metallicity region with upcoming space missions.

\section{Conclusions}
\label{sec:conc}
In this paper, we used a forward-modelling approach to match the width of the RGB bump and the sharpness of the edge formed by ZAHeB stars. Matching the broadening of those features between the \kepler{} and \galaxia{} samples allowed us to constrain the intrinsic scatter of the asteroseismic scaling relations. 

The main results are summarised in Figs.~\ref{fig:limits-rgb} and~\ref{fig:limits-heb}. We found that the observed broadening arises primarily from the measurement uncertainties of \Dnu{} and \numax{}. By taking into account the uncertainty reported by the SYD pipeline, the scaling relations have intrinsic scatter have values of $\sim0.5\%$ (\Dnu{}), $\sim1.1\%$ (\numax{}), $\sim1.7\%$ ($M$) and $\sim0.4\%$ ($R$). This confirms the remarkable constraining power of the scaling relations. The above numbers are appoximate, bacause the systematic uncertainties of our method arising from identifying the features is on a similar level. Although this result was obtained using stars in a limited parameter space, we expect they are applicable to a broader population spanning most low-mass red giants, provided they have similar surface properties.  

Moreover, we demonstrate that using the theoretically corrected \Dnu{} does not reduce the scatter by a large amount. We also found a marginal dependence of the intrinsic scatter of the radius scaling relation on mass and metallicity. However these interpretations are limited by the systematic uncertainties of our method. 

Future work could include using more data from both asteroseismology and spectroscopy to allow tests in more mass and metallicity bins, especially improving the constraints for secondary clump stars. Additionally, by considering the position of those features and matching the exact distributions of stellar parameters (instead of simply the distances to the edge), one could provide constraints on physical processes such as convection and mass loss, and potentially on the offset of the scaling relations.


\section*{Acknowledgements}
We thank the \emph{Kepler} Discovery mission funded by NASA’s Science Mission Directorate for the incredible quality of data. 
We acknowledge funding from the Australian Research Council, and the Joint Research Fund in Astronomy (U2031203)
  under cooperative agreement between the National Natural Science
  Foundation of China (NSFC) and Chinese Academy of Sciences (CAS). 
This work is made possible by the following open-source Python software: {\small Numpy} \citep{numpy}, {\small Scipy} \citep{scipy}, {\small Matplotlib} \citep{matplotlib}, {\small Corner} \citep{corner}, and {\small EMCEE} \citep{emcee}. 

\section*{Data Availability}
The code repository for this work is available on Github.\footnote{https://github.com/parallelpro/nike}
The data sets will be shared on request to the corresponding author.




\bibliographystyle{mnras}
\bibliography{ref/myastrobib.bib} 

\bsp	
\label{lastpage}
\end{document}